\documentclass[%
 reprint,
superscriptaddress,
 amsmath,amssymb,
 aps,
floatfix,
]{revtex4-1}

\usepackage{graphicx}
\usepackage{xcolor} 
\usepackage{enumerate}
\usepackage{dcolumn}
\usepackage{bm}
\usepackage{hyperref}

\begin{document}

\preprint{APS/123-QED}

\title{Interface growth driven by a single active particle}

\author{Prachi Bisht}
 \email{bishtp4496@gmail.com}
\affiliation{%
 TIFR Centre for Interdisciplinary Sciences, Tata Institute of Fundamental Research, Gopanpally, Hyderabad 500107, India
}
\affiliation{%
Indian Institute of Space Science and Technology, Thiruvananthapuram, Kerala 695547, India
}
\author{Mustansir Barma}%
 \email{barma23@gmail.com}
\affiliation{%
 TIFR Centre for Interdisciplinary Sciences, Tata Institute of Fundamental Research, Gopanpally, Hyderabad 500107, India
}%

\date{\today}

\begin{abstract}
We study pattern formation, fluctuations and scaling induced by a growth-promoting
active walker on an otherwise static interface. Active particles on an interface define a
simple model for energy-consuming proteins embedded in the plasma membrane, responsible for
membrane deformation and cell movement. In our model, the active particle overturns local valleys of
the interface into hills, simulating growth, while itself sliding and seeking new valleys. In
1D, this “overturn-slide-search” dynamics of the active particle causes it to move
superdiffusively in the transverse direction while pulling the immobile interface upwards. Using Monte Carlo simulations, we find an emerging tent-like mean profile developing with time, despite large fluctuations. The roughness of the interface follows scaling with the growth, dynamic, and roughness exponents, derived using simple arguments as $\beta=2/3, z=3/2, \alpha=1/2$ respectively, implying a breakdown of the usual scaling law $\beta = \alpha/z$, owing to very local growth of the interface. The transverse displacement of the puller on the interface scales as $\sim t^{2/3}$ and the probability distribution of its displacement is bimodal, with an unusual linear cusp at the origin. Both the mean interface pattern and probability distribution display scaling. A puller on a static 2D interface also displays aspects of scaling in the mean profile and	probability distribution. We also show that a pusher on a fluctuating interface moves subdiffusively leading to a separation of time scale in pusher motion and interface response.

\end{abstract}

\maketitle


\section{Introduction}
\label{intro}

    Active particles are agents that consume energy from a replenishable energy source and thereby propel themselves or generate mechanical forces on other bodies \cite{activematter}. When coupled to a stationary, pliable interface, such particles exert non-thermal forces that can drive the interface into motion, and significantly modify the interface shape and morphology \cite{prost96,madan01} while themselves displaying interesting self-organization, even if they do not interact directly and their motion is dictated only by the local environment \cite{act2}.

\begin{figure}[ht!]
    	\centering
    	\includegraphics[scale=1.0]{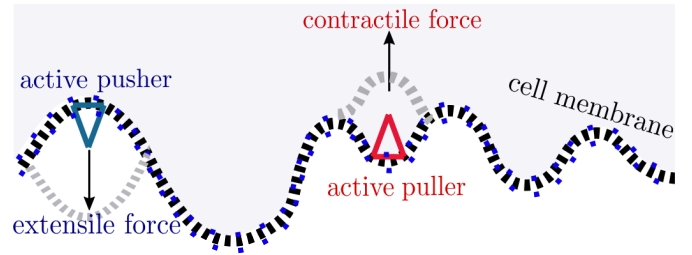}
    	\caption{\label{fig:interface} Cell membrane modelled as a flexible 1D interface with active proteins studded on it. The extensile and contractile forces exerted by the cytoskeleton are mediated by the transmembrane components, and they result, respectively, in a pushing (blue) or pulling (red) effect on the interface.}
    \end{figure}

    The study of active agents on an interface derives motivation from a natural biological setting, namely the interaction of active proteins with the plasma membrane of a living cell. Transmembrane proteins, which derive their energy from the cytoskeleton, are known to exert active forces on the membrane; these forces can be extensile or contractile in nature (Fig. \ref{fig:interface}), and they are consequential in attaining a robust shape deformation of the membrane and for cell migration \cite{fluidmos,bourne02,pollard03,gov18}. More generally, forces and flows generated by motor proteins bring in new length scales into the system and provide a basis for membrane patterning and morphogenesis \cite{howard}.

\begin{figure}[ht]
    \centering
    \includegraphics[scale = 0.4]{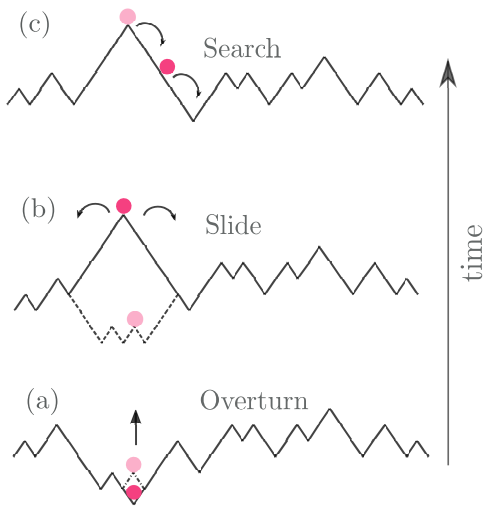}
    \caption{\label{puller_dyn}Dynamical moves of a puller performing an ``overturn-slide-search" motion on an interface. (a) When in a local valley the puller overturns it to a hill, inducing interface growth. (b) Once it is on top of a local hill, the particle chooses at random a direction along which it will slide down, (c) searching, until it finds a local valley, which it then overturns. }
    \end{figure}

    In this paper, we study a minimal statistical physics model that encapsulates some features pertaining to patterning on an active membrane. Specifically, we study the simplest case: that of a \textit{single} active particle interacting with an inert interface initially at rest. Interesting effects arise when such a particle is allowed freely to move on a static interface that is susceptible to deformations by the particle; in turn, the interface profile influences the sliding motion of the particle in a randomly chosen direction. The dynamics of the particle incorporates elements of stochasticity and search in a medium that is itself modified by the trajectory of the particle (Fig. \ref{puller_dyn}). The resulting profile exhibits power
    laws in space and time, broadly reminiscent of the self-organized critical state induced by an Eulerian
    walker that modifies the properties of the background as it moves along \cite{priezzhev96}. The
    simplicity of our model lends it a significant advantage: it allows a detailed understanding of pattern	formation, fluctuations and scaling properties of active membranes.
    
    To start with, we study the interaction of an active particle that pulls the interface locally, hence it is referred to as a ``puller"; on the other hand, a ``pusher" pushes the interface locally. Our study is closely related to that of Cagnetta et al. \cite{cagnetta19} on the dynamics
    of an active growth-promoting slider (puller) on a fluctuating interface. Our work emphasizes the case
    of a \textit{static} interface in one and two dimensions, and it characterizes scaling properties of large-scale structures and probability distributions, for both pullers and pushers.

    The dynamics of a puller gives rise to an interesting walk on the interface, which is composed of local
    slopes, hills, and valleys (see Fig. \ref{puller_dyn}). If found in a valley, the walker overturns it into a hill,
    causing the overall interface height to rise slightly (hence the name puller). Following the overturn, the
    particle finds itself on top of a hill and slides down a randomly chosen direction till it finds a valley to
    overturn and the sequence repeats on the modified interface. With its disposition to convert valleys into
    hills, the ``overturn-slide-search" sequence leads to an ever-evolving walk in an infinite
    system.

    The dynamics gives rise to superdiffusive transverse motion for the puller with the probability of returning to the origin being smaller than for a Brownian random walk. Alongside, the succession of valley to hill conversions results in interesting morphological changes of the surface on a large scale: the mean profile shows a distinctive power-law pattern despite strong fluctuations in individual evolutions. This pattern is sensitive to macroscopic changes of the initial conditions and also is very different for a pusher.   
   
    
    \subsection{Earlier related work}
    \label{earlierwork}
    We briefly review earlier related work on models of active biomembranes, emphasizing the relationship with our work.
     
    The active character of a biomembrane comes from proteins that act as a source of force applied ``inwards" pointing normally towards the interior of the cell or ``outwards" towards the exterior (Fig. \ref{fig:interface}) after seeking out regions whose curvature is compatible with their structure. Continuum field theories that incorporate these effects were studied in \cite{act2,gopi06,veksler07}. For the two functionally different kinds of active protein, two qualitatively different membrane morphologies and protein organization emerge: a phase in which protein density fluctuations travel as waves, and another in which the fluctuations undergo coarsening leading to clumping of proteins in small regions. In our model, active particles seek local valleys. This is akin to, though different in detail from, proteins seeking shape-conforming curvature as ``hot-spots" where active forces act. With a macroscopic number of particles, we too find two phases, as will be discussed elsewhere \cite{unpub}.
    
    A coupled non-equilibrium system of two types of active, hard-core particles on a fluctuating landscape was studied by Chakraborty et al. \cite{shauri2017a,shauri2017b}, inspired in part by the two-way interaction between cell interface components and the cytoskeletal cortex. Tuning the differential activity between the two species leads to a transition from a disordered phase that supports kinematic waves to several sorts of ordered phases. In the latter, particle species are phase-separated in all cases, but interface morphologies vary. In the special case in which there is a single particle of one species, this model reduces to the active particle model under study in this paper.
    
   Our work is closely related to that of Cagnetta et al. \cite{cagnetta18,cagnetta19} who studied the dynamics of active, non-interacting growth-promoting particles on a fluctuating interface. They found that proteins organize in micro-clusters and drive the fluctuating front, mimicking a migrating cell membrane, while themselves self-organizing in transverse traveling waves. In a detailed study of a single growth-promoting active puller on a fluctuating interface, they studied the effect of $\omega$, the ratio of interface to particle update speed \cite{cagnetta19}. When $\omega\gg1$, an interesting surfing regime emerges, where the active particle rides ballistically on a wave created by itself. Our model differs from \cite{cagnetta19} as our static interface cannot be attained by taking the limit $\omega \rightarrow 0$. The static interface is not adiabatic in the strict sense; while it has no underlying fluctuations of its own, it is amenable to deformation by the active particle.
    
    Phase transitions that ensue with a macroscopic number of pullers/pushers on a static or fluctuating interface will be discussed elsewhere \cite{unpub}.


    \subsection{Summary of results}
    \label{sum}

    \begin{figure}[ht!]
        \hspace{-1cm}
        \includegraphics[scale = 0.65]{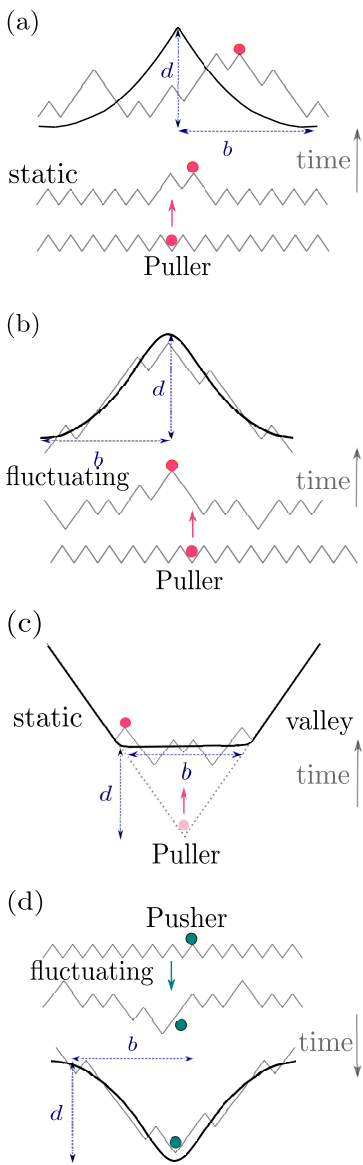}
        \caption{\label{fig:123}(a) Puller on a static, initially ``flat", interface. The particle pulls the interface upwards as it slides over it. The black solid line represents the profile  averaged over many realization, at a given time. Profiles in individual evolutions show large deviations from the mean. (b) Puller on a fluctuating interface constructing a mean mountain-like profile. (c) A puller that started out from the bottom of a macroscopic valley, riding on a rough expanding front. (d) Pusher on a fluctuating interface giving rise to a mean valley like profile. }
    \end{figure}
    
   In this subsection, we discuss the principal results of this paper in brief:
    
    \noindent \textit{Puller on a static one-dimensional (1D) interface}:  Initially, a single puller is placed at the origin on a flat, static, interface. It slides down towards valleys and flips them to hills. The coupled particle-interface motion has interesting consequences:
    \begin{itemize}
        \item A distinctive tent-shaped pattern of the interface profile $H(x,t)$ emerges (Fig. \ref{fig:123}a) characterized by a base length $b$ and height $d$, growing in time as $b\sim t^{\lambda}$ and $d\sim t^{\mu}$. We observe $\lambda + \mu = 1$ and $\lambda = 1/z$, where $z=3/2$, is the dynamic exponent. Fluctuations of the height are of the same order of magnitude as the mean height itself (both $\sim t^{1/3}$).
        \item The lateral motion of the active particle is superdiffusive, and its rms displacement, $\sqrt{<x^2(t)>}$, grows as $\sim t^{\lambda}$, similar to the base of the tent. The corresponding probability distribution $P(x,t)$ is bimodal with a linear cusp at the origin. $P(x,t)$ is a scaling function with a horizontal length scale growing as $\sim t^{\lambda}$.
        \item If the system size $L$ is finite, the steady state is reached in time $\tau_L \sim O(L^z)$ with $z=3/2$. For $t\gg \tau_L$, the interface rises upwards with a constant speed proportional to $1/L$, and it is characterized by a roughness that grows as $\sim L^{\alpha}$, where $\alpha = 1/2$.
        \item In the early time regime ($t\ll \tau_L$), the width increases as $\sim t^{\beta}$, where $\beta$ is the growth exponent and takes the value $2/3$. This value is anomalous in that the familiar scaling relation $z=\alpha/\beta$ fails. The failure is traced to the fact that the height at only a single site is updated in each Monte Carlo step; this brings in $L$-dependent factors in the early time scaling. The values of the exponents $z,\alpha$ and $\beta$ are derived using simple arguments based on plausible assumptions and a revised scaling law is proposed, verified by numerical simulations.
    \end{itemize}

 \noindent {\textit{Puller on a fluctuating 1D interface}}: This regime, where a single puller is coupled to an Edwards-Wilkinson (EW) interface \cite{ew}, was studied in considerable depth in \cite{cagnetta19}. We observe a mean pattern $H(x,t)$ in the interface profile in this regime as well. Both $H(x,t)$ and the probability distribution $P(x,t)$ of the particle obey similar scaling laws as for the puller on a static interface. The profile has a smooth maximum (Fig. \ref{fig:123}b), and $P(x,t)$ shows a rounded dip instead of a cusp at the origin.\\

\noindent {\textit{Puller in a macroscopic valley}}: A macroscopically different initial condition of the interface strongly influences the pattern formation, the motion of the particle and height fluctuations. This is illustrated by imposing a cutoff on the maximum transverse displacement of the particle, in the form of a macroscopic valley, as shown in Fig. \ref{fig:123}(c), with the puller initially at the bottom. In time we see the particle rides on a rough upward-advancing front. The effect of this interface geometry is to limit the growth of displacement and base length $b$ of the front to a rate $\sim t^{1/2}$ implying $\lambda = 1/2$. Also the height grows as $d \sim t^{1/2}$, again satisfying the condition $\lambda + \mu = 1$. Mean-square fluctuations in height are of order mean height and grow as $t^{1/2}$ as well. \\

\noindent {\textit{Pusher on a fluctuating 1D interface}}: Like a puller, a pusher slides towards valleys; but unlike a puller, the pusher has a preferential tendency to overturn hills into valleys (Fig. \ref{fig:123}d), giving rise to a mean profile in the form of a valley about the pusher's initial location. The pusher motion is observed to be subdiffusive due to its ``digging" activity on the interface. We again observe scaling in $H(x,t)$ and $P(x,t)$. However, unlike the puller on a fluctuating interface, the displacement of the pusher and profile base length have different power-law growths. \\

\begin{figure}[ht]
    \includegraphics[scale = 0.45]{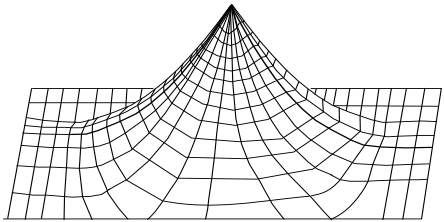}
    \caption{\label{fig:2dtent}Puller on a static, flat 2D interface gives rise to a circular tent-like structure analogous to the profile seen in one dimension.}
   \end{figure}

\noindent  {\textit{Puller on a static 2D interface}}: For an active puller on a 2D static interface, we observe scaling features in observables like the mean profile (Fig. \ref{fig:2dtent}) and the probability distribution for particle position. Both are circularly symmetric about the origin. The scaling exponents here are $\lambda \simeq 0.53$  and $\mu \simeq 0.46$.

\section{Model and Parameter Space}
\label{model}


\subsection{One Dimension}

In one dimension, the model consists of a flexible lattice of length $L$ and spacing $a$ with a periodic boundary condition. Each site in the lattice is linked with a bond placed at a half-integer site that takes values $\tau_{x+1/2} = \pm 1$, where $x$ runs over integer values.  The height of a site is given by $h_y = \sum^y_x \tau_{x+1/2}$ and follows the solid-on-solid restriction $|h_{x+1} - h_x| = 1$. The periodic boundary condition $h_N = h_0$ implies $\sum^N_0 {\tau_{x+1/2}}=0$. The active particle occupies the integer sites between bonds.

\noindent  \textit{Interface Update:} The interface is evolved via stochastic local single step moves, i.e in an infinitesimal time interval $dt$ at most one update is performed and it only occurs if the randomly chosen site is on a local hill or a valley. A hill (valley) flips to a valley (hill) with probability $p_+dt$ ($p_-dt$) and is accompanied by a change in height $h_x$ of the site of the hill (valley) by $-2a$  $(+2a)$. The rate of valley to hill overturn ($\vee \rightarrow \wedge$) and hill to valley overturn ($\wedge \rightarrow \vee$) depends on whether the chosen site holds the particle or not, as discussed below.
    
        \begin{enumerate}
            \item When no particle is present, the chosen valley (hill) overturns into a hill (valley) with a rate $u/2$ (Fig. \ref{fig:surface_updates}a). Thus we have local Edwards-Wilkinson (EW) moves at all sites devoid of the particle.  This move represents the fluctuations inherent to the interface and is characterized by rate $u$. For a static interface, we have $u=0$, a case that is of special interest to us.
            
            \item When the active particle is at the chosen site, an update follows with the rates, $p_+$ and $p_-$, given by:
            
            \begin{equation}
                p_+ = w\frac{1}{1+e^{-2\beta_o}}
                \label{p_+}
            \end{equation}

            \begin{equation}
                p_- = w\frac{e^{-2\beta_o}}{1+e^{-2\beta_o}}
                \label{p_-}
            \end{equation}
             
             where $\beta_o$ is the activity parameter.    
         
            \begin{figure}[ht!]
            \centering
            \includegraphics[scale = 0.5]{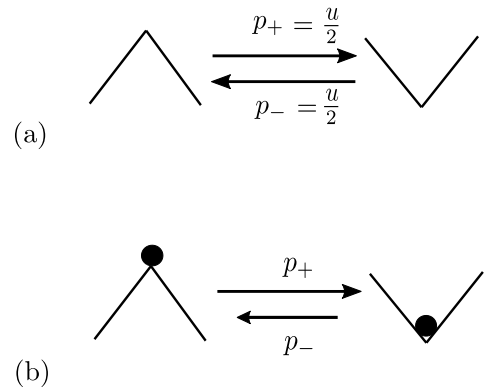}
            \caption{\label{fig:surface_updates}(a) Probability rates when no particle is present. (b) Probability rates when a particle is present. If $\beta_o>0$, we have $p_+>p_-$.}
            \end{figure}

            For $\beta_o > 0$, we have $p_+ > p_-$, that implies that the transition of a hill laden with the particle overturning into a valley is more likely than the reverse. In this regime, the active particle acts as a pusher. For  $\beta_o < 0$, we have $p_- > p_+$; there is a higher likelihood for valleys with particles overturning into hills, making the particle a puller in this regime.
         \end{enumerate}

 Evidently, interface evolution follows EW dynamics around sites that hold no particle ($p_+ = p_- = u/2$). At the site that holds the active particle the symmetry between $p_+$ and $p_-$ is broken giving rise to local KPZ like dynamics.

\begin{figure}[ht!]
    \centering
    \includegraphics[scale=0.75]{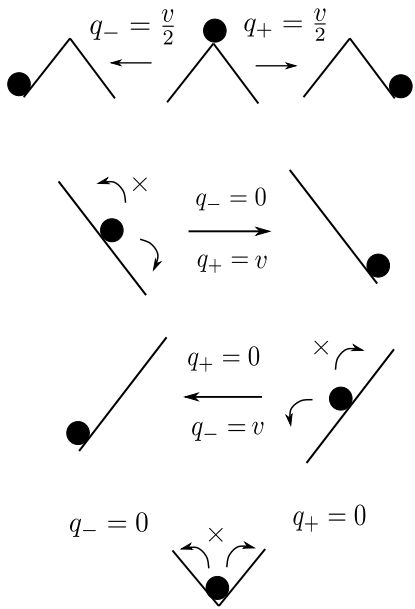}
    \caption{The particle diffuses on either side when on top of a hill, drifts downward when on a slope and faces a barrier to jump, when in a valley.}
    \label{fig:pcle_update}
\end{figure}

\noindent \textit{Particle Update:} An interface update is followed by selecting the particle with probability $1/L$ and allowing it to hop one lattice site away as shown in Fig. \ref{fig:pcle_update}. The interface poses as a potential landscape over which the particle tends to slide towards a local minimum. If $q_-$ and $q_+$ are the left and right hopping probability rates, both are equal to $v/2$ if the particle is found on top of a local hill i.e. the particle will slide down in a direction randomly chosen between left and right. If the particle is found on a slope, it will slide down until it finds the nearest valley. And if found in a local valley, the particle cannot escape unless there is a valley to hill transition at the site of the particle. Figure \ref{fig:pcle_update} illustrates the probability rates for all possible moves.

To summarize, the active particle always slides towards the local minimum of the interface. It overturns valleys to hills and induces an upward (downward) movement of the interface if it is a puller (pusher).

Each interface update is followed by a particle update, and the pair constitutes a micro-step. Further, $L$ micro-steps constitute 1 Monte-Carlo (MC) step, the unit of time. On an average, in 1 MC, each interface site and the particle are accessed once.  For our purposes the interface to particle update ratio is always 1.

\begin{figure}[ht!]
    \hspace{-1cm}
    \includegraphics[scale=0.3]{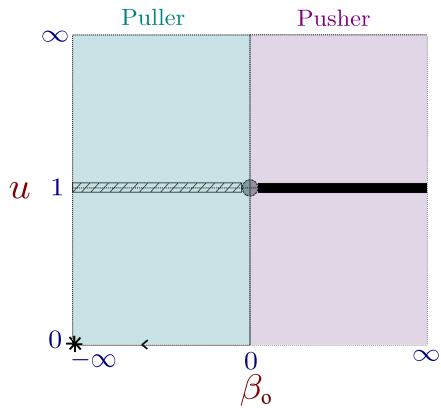}
    \caption{\label{1p_phasespace} The parameter space for our model. The parameter $u$ represents the rate of intrinsic fluctuations of the interface while $\beta_o$ is the activity parameter. The left half: the $\beta_o < 0$ (puller) regime was studied in \cite{cagnetta19}. The $u=1$ line corresponds to their $\omega = 1$ regime. The focus of our paper is on the regimes $\beta_o < 0, u=0$ (puller) and $\beta_o>0, u=1$ (pusher).}
    \label{fig:1p_phasespace}
\end{figure}

In our work we set $w,v=1$, while $u$ and $\beta_o$ govern the degree of intrinsic and active interface fluctuations respectively, forming the axes of our dynamic parameter space. (Fig. \ref{fig:1p_phasespace}). The left half ($\beta_o <0$) represents the puller and the right half ($\beta_o >0$) represents the pusher. The point with the coordinates $u = 1, \beta_o = 0$ is the case of a passive slider on a fluctuating interface, that has been studied extensively \cite{chin02,drossel00,manoj04,nagar06,singha18}.

As mentioned earlier, setting $u=0$ makes the interface static.  This implies that a site devoid of the active particle, if chosen, undergoes no update. If the site with the particle is chosen, it is updated as per Eqs. (\ref{p_+}) and (\ref{p_-}). To save computational time, we modify the evolution algorithm such that in one MC only the site containing the particle undergoes an update, followed by a certain update of the particle as per the rules shown in Fig. \ref{fig:pcle_update}. We have checked that the qualitative results remain the same with this change, with the only quantitative difference arising due to a rescaling of time. 


    \subsection{Two dimensions}

    The interface in 2D is in the form of a square grid where each site follows the solid-on-solid condition, i.e., the height difference between nearest neighbors is maintained at $\pm1$. The interface follows the periodic boundary condition in both directions. The active particle is placed initially at the origin.
 
    For the active puller on a static interface, the update algorithm is as follows:
    
    \noindent \textbf{Interface update}: The site with the puller undergoes an update in height by $+2$ only if it is lower in height than all of its four neighbors. Otherwise no update occurs \cite{meakin,manoj2d}.
    
    \noindent \textbf{Particle update}: Following a interface update, the particle randomly chooses one of the four nearest-neighbor sites. If the height of the chosen site is lower than that of the current site, the particle hops to the chosen site. Otherwise there is no update.\\
    \noindent An interface update followed by a particle update constitutes a Monte-Carlo step.
    
    Due to the discrete nature of the lattice both in 1D and 2D, we cannot get a perfectly flat interface. A flat interface in our model looks like the jagged structure shown in Fig. (\ref{fig:123}), where an alternate site has a height of 0 and the rest have a height of 1. In our analysis of a ``flat" interface, we record observations only at the alternate sites with an initial height of 0.


    \section{Puller on a static 1d interface}
    \label{pull_stat}

    Consider a puller starting from the origin and moving on a flat, static interface $(u=0$) through overturn-slide-search dynamics defined in Sec. \ref{intro}. For simplicity, we take $\beta_o \rightarrow -\infty$ in Eqs. \ref{p_+} and \ref{p_-}, implying that only $\vee \rightarrow \wedge$ moves are permitted at the site of the particle.

    \noindent In Sec. \ref{interface profile and fluctuations}, we study pattern formation in the ensemble-averaged profile and fluctuations about it. The roughness of the interface and the scaling properties are discussed in Sec. \ref{Scaling Properties}. A breakdown of the customary scaling relation linking the growth, dynamic, and roughness exponents is observed. Section \ref{Particle motion} deals with the transverse motion of the puller, the scaling properties of its probability distribution in space, and its recurrence properties in time. 
    
    
    \subsection{Interface profile and fluctuations}
    \label{interface profile and fluctuations}
\begin{figure}[ht]
        \centering
        \includegraphics[scale=0.55]{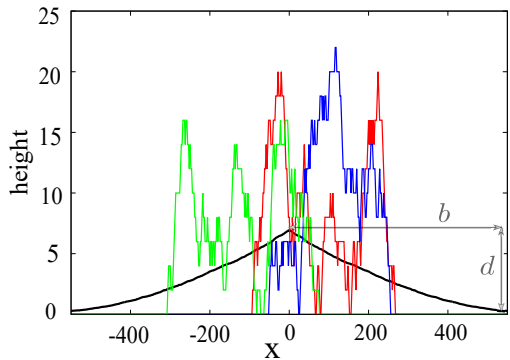}
        \caption{ The colored lines are the height profiles at the end of a different realization at $t=5000$, for a particle starting from $x=0$ at $t=0$. The black tent-like structure is an ensemble-averaged profile at a given time with height $d$ and base length $b$ as depicted. Evidently fluctuations are large.}
        \label{fig:typ0}
    \end{figure}

    As the particle pulls valleys upward, the interface profile builds up in time. Figure \ref{fig:typ0} shows that the particle sculpts large structures that vary considerably from history to history. The ensemble-averaged profile, however, shows a distinctive pattern fixed in space with a tent-like structure centered about the origin (Fig. \ref{fig:typ0}).
    
\begin{figure}[ht]
        \centering
        \includegraphics[scale=0.56]{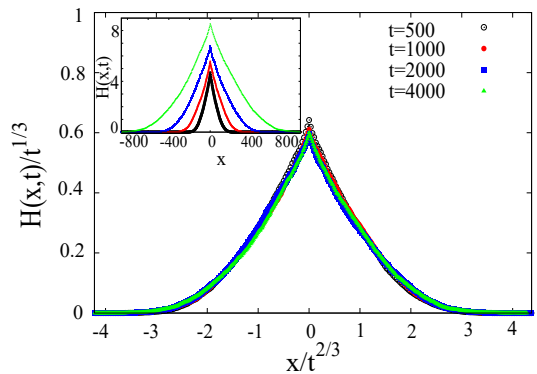}
        \caption{The scaled form of the mean interface profile constructed by the puller, obtained after averaging over an ensemble of histories. The inset shows the unscaled mean profile at different times.}
        \label{fig:profile}
    \end{figure}

    The height $d$ and base length $b$ of the tent grow with time as $\sim t^{\mu}$ and $\sim t^{\lambda}$ respectively. The area under the tent, $A$, is proportional to $b \times d$ and hence grows like $t^{\lambda+\mu}$. It is observed that $\lambda$ and $\mu$ satisfy $ \lambda + \mu = 1$. This follows from the fact that in the course of overturn-slide-search dynamics , the particle spends a finite fraction of the walk time in local valleys, which it promptly overturns; each such overturn increases the area by one unit. Therefore, the area under the risen region grows as $A\approx kt$, implying $\lambda + \mu = 1$. We observe that $k\simeq 0.66$ while the exponents $\lambda$ and $\mu$ are close to $\frac{2}{3}$ and $\frac{1}{3}$, respectively. The mean profile follows the scaling form:

    \begin{equation}
        H(x,t) \approx t^{\mu}\mathcal{H}(\frac{x}{t^{\lambda}})
        \label{profile_scaling}
    \end{equation}
    as evidenced by the scaling collapse shown in Fig. \ref{fig:profile}.
    
   In a single realization however, the risen region of the profile typically overshoots the mean height in magnitude (see Fig. \ref{fig:typ0}) and is positioned predominantly on one side of the origin. To quantify height fluctuations, we monitor the local mean-square deviations $w_{\rm{local}}(x,t)$ away from the mean pattern:
    \begin{equation}
        w_{\rm{local}}^2(x,t) = <(h(x,t) - H(x,t))^2>
    \end{equation}

\begin{figure}[ht]
        \centering
        \includegraphics[scale=0.5]{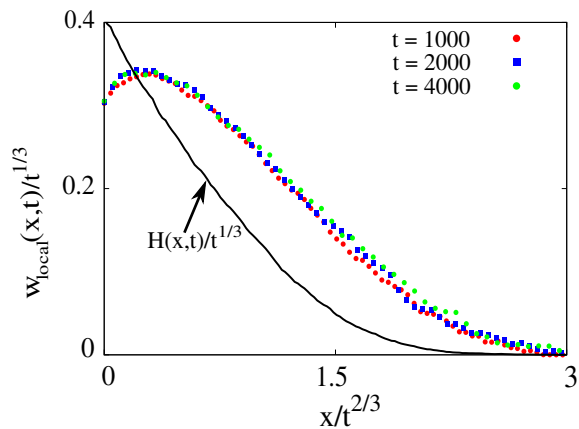}
        \caption{The local roughness in scaled form. The black line shows the scaled mean profile; the local fluctuations are larger in magnitude than the mean height. }
        \label{fig:rough}
    \end{figure}

    Figure \ref{fig:rough} demonstrates that $w_{\rm{local}}(x,t)$ is of the order of the mean height $H(x,t)$ (both $\sim t^{1/3})$ and displays a scaling behavior similar to the profile height:
   
    \begin{equation}
        w_{\rm{local}}^2(x,t) = t^{2\mu}\mathcal{W}(\frac{x}{t^{\lambda}})
    \end{equation}
    
    \noindent with $\mu = \frac{1}{3}$ and $\lambda = \frac{2}{3}$ (see Fig. \ref{fig:rough}). Thus, the local height fluctuations are of the same order as the mean height.
    
    \subsection{Scaling properties}
     \label{Scaling Properties}
     
    We turn to a discussion of scaling properties of height fluctuations in a large but finite system of size $L$. Studies of interface growth models lead us to expect a change of form of dynamical quantities when the time $t$ crosses $\tau_L$, a size-dependent time scale that grows as $\tau_L \sim L^z$ \cite{barabasi}. Here $z$ is the dynamic critical exponent. A useful quantity to monitor is the overall width of the interface,
    
    \begin{equation}
        w^2(t,L) = \frac{1}{L}<\sum_{x=0}^{L}(h(x,t) - \overline{h})^2> 
        \label{fam_vic_eqn}
    \end{equation}
    
    \noindent where $\overline{h} = \frac{1}{L}\sum_{0}^{L}h(x,t)$ and $<>$ denotes an average performed over several realization of interface configurations at that time.
    
    For most interface growth processes, $w(t,L)$ is consistent with the Family-Viscek scaling form \cite{fam_vic}:
    
    \begin{equation}
        w^2(t,L) = L^{2\alpha}g(\frac{t}{L^z})
    \label{roughness_scale}
    \end{equation}
    
    \noindent where $g$ is a scaling function. Typically, at early times, $w(t,L)$ grows independently of the system size as $\sim t^{\beta}$, where $\beta$ is the growth exponent.  The steady state is achieved on a time scale $\tau_L\sim L^z$ beyond which $w(t,L)$ attains a saturation value $\sim L^{\alpha}$ where $\alpha$ is the roughness exponent. For most growth processes, the exponents $\alpha, \beta, z$ obey the scaling relation:
    
    \begin{equation}
        z = \frac{\alpha}{\beta}
        \label{abz_rel}
    \end{equation}
    
    \noindent tying together the early and steady state behavior \cite{fam_vic,barabasi,kpz,ew}. However Eq. (\ref{abz_rel}) does not hold for the pulled static interface in our study.
    
    \noindent We proceed point-wise to argue for the form of relationships between various exponents. 
    \begin{itemize}
        \item For $t\ll \tau_L$, a height profile forms with lateral spread $b\sim t^{\lambda}$ and a vertical spread $d\sim t^{\mu}$ as discussed in Sec. \ref{interface profile and fluctuations}. Since the number of overturned valleys grows linearly with time $t$, we have
        \begin{equation}
            \lambda + \mu = 1
            \label{arealaw}
        \end{equation}{}
        \item For $t\gg \tau_L$, the system reaches steady state, after which the active particle traverses and spans the full system several times. In the process, the height profile gets randomized; we have checked numerically that correlations between local slopes are very short-ranged. Consequently the roughness, quantified by fluctuations from the mean, attains the form $w^2_{steady state} \sim L$. Thus the roughness exponent $\alpha$ has the value:
        \begin{equation}
            \alpha = \frac{1}{2}
            \label{alpha}
        \end{equation}{}
        \item As $t$ increases towards $\tau_L$, the breadth of the profile $b$ approaches $L$. Matching the associated time taken, $L^{1/\lambda}$, with $\tau_L \sim L^z$, we read off $\lambda = 1/z$. Furthermore, the fluctuations in height at that stage vary as $\sim \tau_L^{\mu}$. Comparing with the steady-state roughness $\sim L^{\alpha}$, we obtain,
        \begin{equation}
            \frac{\mu}{\lambda} = \alpha
            \label{mu/lambda}
        \end{equation}{}
        \item Equations (\ref{arealaw}),(\ref{alpha}),(\ref{mu/lambda}) together yield $\lambda = \frac{2}{3}$ and $\mu = \frac{1}{3}$, which we observed in Eq. (\ref{profile_scaling}) in previous section. Recalling that $\lambda = 1/z$, we see that $z=3/2$. Notice that we have been able to deduce the exponents $z$ and $\alpha$ from two simple, plausible assumptions (i) the area under the risen interface growing linearly in time (ii) the randomization of the height profile in steady state. 
        
        \item Finally, let us characterize the early time ($t\ll \tau_L$) growth of width $w(t,L)$. With our microscopic moves, the growth in a time step occurs locally at one site, unlike the uniform stochastic evolution over $L$ sites in typical growth models. Let us assume a power-law growth of width at early times:
        \begin{equation}
            w^2(t,L) \approx c(L) t^{2\beta}
        \end{equation}{}where $c(L)$ is a time-independent constant that may depend on system size.
        Now at $t=1$, we know that a single valley at the origin is flipped into a hill by the puller. It is thus straightforward to evaluate the right-hand side of Eq. (\ref{fam_vic_eqn}). On substituting the average height $\overline{h} = 1/L$ and $h_0 = 1, h_x = 0$ for all $x\neq 0$, we get:
        
        \begin{equation}
            c(L) \approx \frac{1}{L}
            \label{early_prefactor}
        \end{equation}{}
        \item At the crossover time $t \approx \tau_L \sim L^z$, we match the short time $(t < \tau_L$) and the large time ($t>\tau_L$) forms of $w^2$ to obtain:        \begin{equation}
            \frac{1}{L}\tau_L^{2\beta} \sim L^{2\alpha}
            \label{crossover}
        \end{equation}
        
        \end{itemize}{}

    Substituting $\alpha = 1/2$, we get $\beta = 2/3$. Note that this value of $\beta$ marks a breakdown of the scaling relation Eq. (\ref{abz_rel}). This is an outcome of the fact that the puller acts extremely locally, only at a single site at each time step.
        
\begin{figure}[ht!]
       \centering
       \includegraphics[scale=0.34]{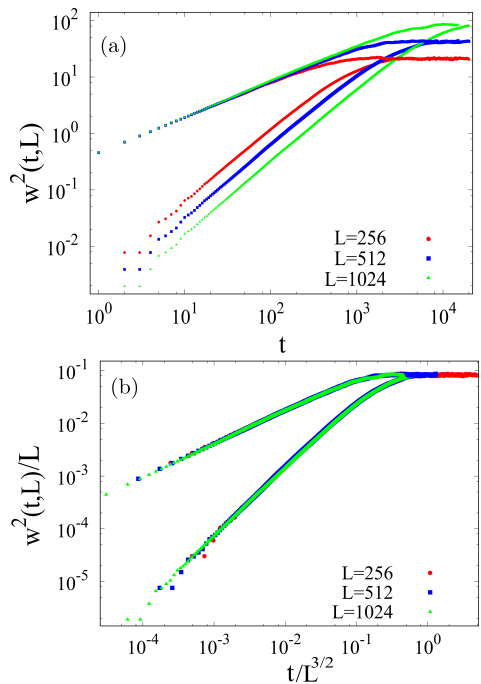}
       \caption{The roughness plot for a pulled static interface (lower) and KPZ interface (upper). The plot in (a) shows the system size dependence of the roughness at early times for the pulled interface, causing the deviation from the relation in Eq. \ref{abz_rel}. The scaled plots in (b) demonstrate that $\alpha = 1/2$ and $z=3/2$, while $\beta = 2/3$ for the pulled interface and $\beta = 1/3$ for the KPZ interface. The roughness in the figures is plotted  after subtracting its value at $t=0$, i.e., $w^2(t=0,L)= 0.25$.}
       \label{fig:fam_vic}
       \end{figure}

    Figure \ref{fig:fam_vic} shows a numerical verification of these results. For comparison we also include the roughness growth of a Kardar-Parisi-Zhang (KPZ) interface, which satisfies simple size-independent early time growth of the width and the scaling relation in Eq. (\ref{abz_rel}) \cite{kpz}.\\
     
\begin{figure}[ht!]
    	\centering
    	\includegraphics[scale=0.6]{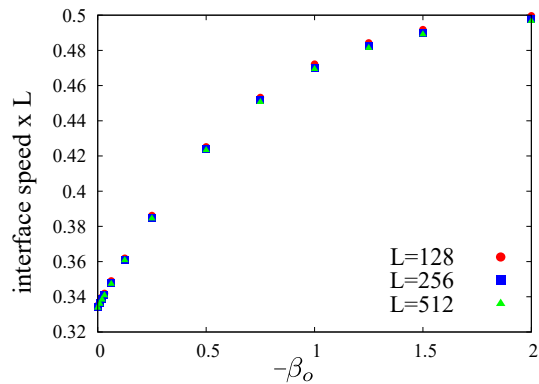}
    	\caption{The interface speed in steady state as $-\beta_o$ is varied. The speed increases linearly for small values of $\beta_o$ until it finally saturates, scaling with interface size as $1/L$. }
    	\label{fig:current_beta_size}
    \end{figure}

    \noindent \textit{Interface motion in steady state}:  For time scales $t\gg L^{z}$ i.e. in steady state, the interface moves upwards with a constant speed. Since the evolution of the interface involves flipping a single valley into a hill and advancing one unit upwards at one site in a time step, the increase of mean height, and thus the overall speed of an interface of size $L$, varies as $\sim 1/L$. Figure \ref{fig:current_beta_size} shows the variation of speed with $\beta_o$.

    Interestingly, we observe that the interface has a non-zero speed of growth even when $\beta_o$ is zero i.e. the particle does not overtly affect the interface (in the passive limit). This feature is a consequence of the update algorithm. For a frozen interface, the only update move possible is at the site of the particle. Since the particle tends to slide towards valleys, the likelihood of finding it in a valley is higher than finding it on a hill. Consequently, there are more $\vee \rightarrow \wedge$ transitions than $\wedge \rightarrow \vee$ transitions at the position of the particle. This leads to a net speed upwards.

    
    \subsection{Particle motion}
    \label{Particle motion}
    
    The walk performed by the active particle is strongly non-Markovian as the particle alters the landscape on the trail it traverses. Upon returning to the origin, the particle faces a completely different landscape to walk on.
    
\begin{figure}[ht!]
    \centering
    \includegraphics[scale=0.315]{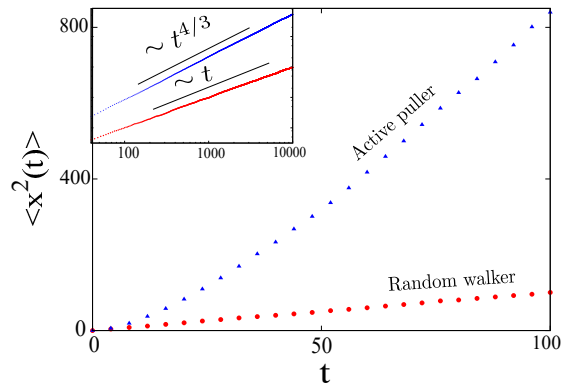}
    \caption{Mean-square displacement of the active particle with reference to a random walker. The motion is faster than diffusive.}
    \label{fig:stats}
    \end{figure}

    The resulting motion of the active particle is superdiffusive. Recognizing that the height profile $H(x,t)$ with base growing as $b \sim t^{\lambda}$ is established by the active particle motion, we conclude that the transverse mean-square displacement of the particle grows as $<x^2(t)> \sim t^{2\lambda} = t^{4/3}$, as shown in Fig. \ref{fig:stats}, where a comparison is made with a random walk.
    
\begin{figure}[ht]
        \centering    
        \includegraphics[scale=0.61]{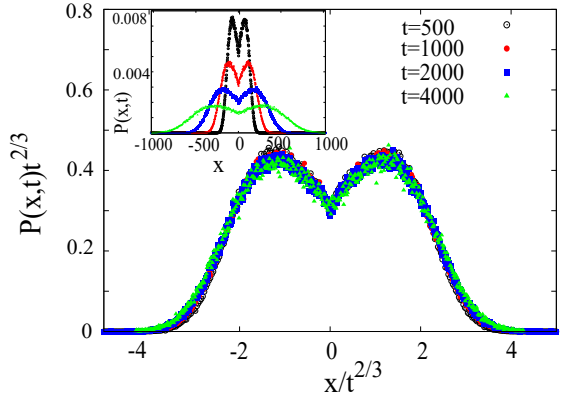}
        \caption{Scaled $P(x,t)$ for the puller on a static interface. Inset: unscaled plots show how $P(x,t)$ spreads with time. }
        \label{fig:prob}
    \end{figure}

	The probability distribution $P(x,t)$ of finding the particle at location $x$ at time $t$, given that it starts
	at the origin at $t=0$, has an interesting form, as shown in Fig. \ref{fig:prob}. With time, the distribution
	spreads in space but with two distinct peaks and a pronounced linear cusp at the origin. The two halves
	of $P(x,t)$, as seen in the inset of Fig. \ref{fig:prob}, move outward as $\sim t^{\lambda}$, where
	$\lambda = 2/3$ as for the growth of the base length $b$. This behavior persists as
	long as the displacement $x$ is much smaller than the system size $L$, or equivalently $t \ll \tau_L \sim
	L^{1/\lambda}$. For $t \gg \tau_L$, the system reaches steady state; the form of the probability
	distribution and the mean-square displacement in this regime are discussed later (see Eq. (18)).

    As evidenced from the scaling collapse shown in Fig. \ref{fig:prob}, for times $t$
    satisfying $t\ll\tau_L$, the probability distribution $P(x,t)$ follows the scaling form:

    \begin{equation}
        P(x,t) \approx t^{-\lambda}\mathcal{P}(\frac{x}{t^{\lambda}})
         \label{prob_scaling}
    \end{equation}
    
    The superdiffusive nature of the particle motion can be traced to its predilection to move along the direction chosen at the first time step, once it forms a hill and slides away from there. This is consistent with the maxima of $P(x,t)$ migrating away from the origin as time increases.
    
\begin{figure}[ht!]
        \centering
        \includegraphics[scale=0.6]{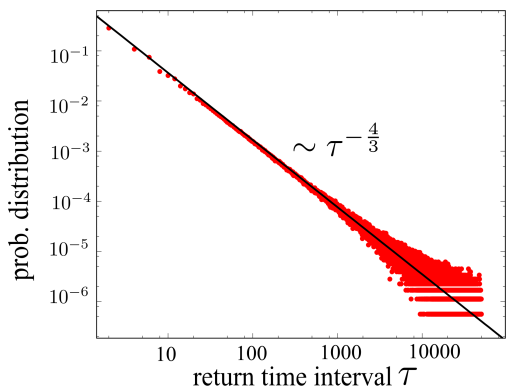}
        \caption{ Probability distribution of return time intervals $\tau$ recorded until some finite time (here 50000 MC). }
        \label{fig:delt_dist}
    \end{figure}

   To quantify how often the particle visits the origin, we study the probability distribution of the return times $\tau$ i.e. the time interval between two consecutive visits to the origin. The probability distribution $P(\tau)$ follows a power law, $P(\tau) \sim 1/\tau^{\theta}$ where $\theta = 4/3$ (Fig. \ref{fig:delt_dist}). For a random walk, we have $\theta=3/2$. The lower value of $\theta$ in our case signifies that the probability of longer return time intervals is enhanced, consistent with the superdiffusive motion of the particle. It is observed numerically that the average number of visits to the origin grows as $\sim t^{1/3}$ with time, the same growth law followed by the mean height at the origin.
   
    \noindent \textit{Relation between height and probability density}:
    Since the growth of the height profile $H(x,t)$ is a consequence of active pulling action, it is natural to ask how $H(x,t)$ is related to the probability distribution $P(x,t)$ of the puller. Let us discuss some qualitative aspects. The initial growth at the origin cannot be sustained continuously as the height of the highest point on the profile cannot support a valley overturn, necessary for growth. The puller constructs a hill, slides downward till it reaches a valley to overturn, and returns to the origin only when other sites have risen higher, allowing it to revisit. The interplay of turning valleys into hills and then sliding away from these hills gives rise to the distinctive forms of $P(x,t)$ and $H(x,t)$. In particular, the point $x=0$ is a local maximum of $H$ but a local minimum of $P$.\\
    
    Let us ask for the expected total number of visits $N(x,t) = <n(x,t)>$ to a site $x$ till time $t$.
    It can easily be understood that $N(x,t)$, is the cumulative probability density till time $t$ i.e. $N(x,t) = \sum_{t'\leq t} P(x,t')$. The average height increment at a given $x$ is proportional to the number of times it has been visited by the puller i.e $H(x,t) = c_oN(x,t)$. It has been observed numerically that the prefactor $c_o$ has a value $\simeq 0.66$. Thus the mean height at a given site at time $t$ is proportional to the cumulative probability density till $t$.
    
    \begin{equation}
         H(x,t) =\int_{0}^{t}c_oP(x,t')dt'
         \label{height_prob_rel}
    \end{equation}
    
\begin{figure}[ht]
        \hspace{-1cm}
        \includegraphics[scale=.59]{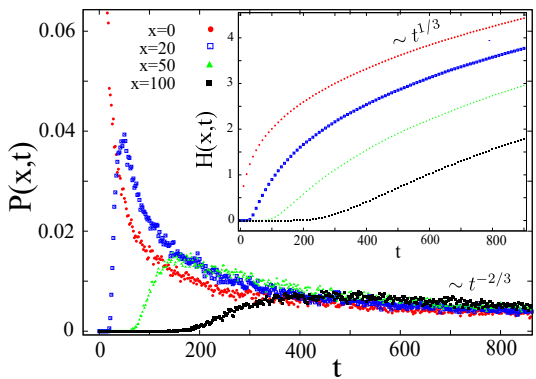}
        \caption{Showing how the probability $P(x,t)$ of being at a given site $x$ evolves with time $t$. In view of Eq. (\ref{height_prob_rel}), the area under the probability decay curve for a given site until time $t$ is a measure of the mean height for that site at $t$. The area under $P(x,t)$ until $t$ being the largest for $x=0$ suggests that $H(x,t)$ at $x=0$ is largest at $t$, as shown in the inset. }
        \label{fig:fixed}
    \end{figure}

    \noindent For the simple case of $x=0$ we get a relation 
    \begin{equation}
         H(0,t) \sim \int_{0}^{t}\frac{1}{t'^{2/3}}\rm{d}t'
    \end{equation}
    giving $H(0,t) = d  \sim t^{1/3}$ as seen in Figs. (\ref{fig:profile}) and (\ref{fig:fixed}). For non-zero $x$, we have $  H(x,t) = \int_{0}^{t}t'^{-\frac{2}{3}}\mathcal{P}(\frac{x}{t'^{\frac{2}{3}}})\rm{d}t'$. 
    At large times i.e. $t\rightarrow \infty$,  we have $\mathcal{P}(\frac{x}{t^{\frac{2}{3}}}) \rightarrow \mathcal{P}(0)$ that is a constant. So, the height growth at any site $x$ should grow as $\sim t^{1/3}$ at large times. 
    Fig. \ref{fig:fixed} shows the $\sim t^{-2/3}$ decay of probability for different values of $x$ and the consequent $\sim t^{1/3}$ form of height growth in the inset.

   \noindent Note that at time $t>0$, the most probable value of the position for the puller lies away from the origin and grows with time as $\sim t^{2/3}$ (refer Fig. \ref{fig:prob}). Despite this, the height profile $H(x,t)$ is maximum at $x=0$ as the height involves cumulative probability $\sum_{t'\leq t} P(x,t')$ or equivalently the total number of visits, which is largest at the origin.\\

    \noindent \textit{Scaling of RMS displacement}: In the discussion so far, we have implicitly assumed that the system size $L$ is large enough that the finite-size effects are not felt in the observations. For instance , the form of $P(x,t)$ in Eq. (\ref{prob_scaling}) holds for $t\ll \tau_L$ where $\tau_L \sim L^{z_p}$ is the typical time for the particle to span the system size $L$. Extending Eq. (\ref{prob_scaling}) to $t\sim \tau_L$, we see that $\tau_L \sim L^{1/\lambda}$, implying $z_p = 1/\lambda = 3/2$. 
    
    Since we have periodic boundary conditions, the displacement $x$ is unbounded and
    can exceed $L$; equivalently on a 1D ring, the particle can go many times around the ring in any
    direction. For large times $t \gg \tau_L$, the probability distribution on the ring approaches
    $P_{\rm{steady}} = 1/L $. However, the mean-square displacement, $r^2(t) = <x^2(t)> $ grows
    with $t$, and we track its behavior in different dynamical regimes. We expect $r^2(t)$ to obey the
    scaling form:

    \begin{equation}
        r^2(t) = L^{2\chi}\mathcal{R}_2(\frac{t}{L^{z_p}})
        \label{p_scale}
    \end{equation}

     \noindent where $z_p$ is the dynamic exponent associated with the puller motion, and the exponent $\chi$ is determined below.
     
     For times $t\ll L^{z_p}$, let us suppose $\mathcal{R}_2(y) \sim y^{\phi}$. The requirement that $r^2(t)$ be independent of $L$ in the early time regime yields $2\chi = z_p\phi$. Further, Eq. (\ref{prob_scaling}) implies $r^2(t) \sim t^{2\lambda}$, which leads to the identification $\phi = 2\lambda = 2/z_p$ with $z_p = 3/2$. Thus we have $\chi = 1$.
     
\begin{figure}[ht!]
        \centering
        \includegraphics[scale=0.3]{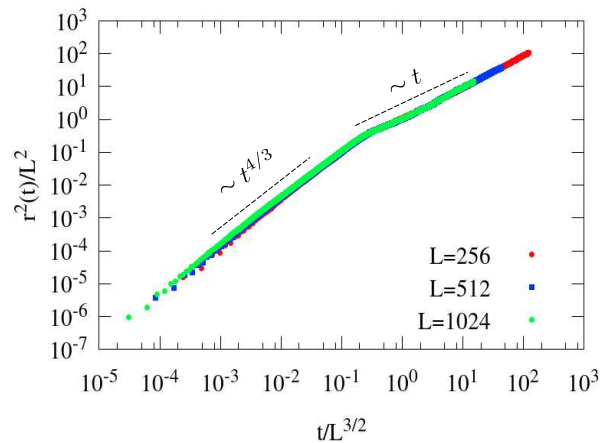}
        \caption{ Scaled mean square displacement. At large times $t \gg L^{z_{p}}$, the walker is free and uncorrelated. The MSD takes the form $r^2(t) \approx D_L t$ in steady state. For early times $t \ll L^{z_{p}}$, the motion of the particle is superdiffusive,  $r^2(t) \sim t^{2\lambda}$, and is independent of system size.}
        \label{fig:rms_scale}
    \end{figure}

     For large times  $t\gg L^{z_p}$, we expect $r^2(t)$ to grow diffusively $\approx
     D_Lt$ with an $L$-dependent diffusion constant. This implies $\mathcal{R}_2(y)$ for large values of $y$,
     leading to $r^2(t) \sim L^{2-z_p}t$. Thus we read off $D_L \sim L^{2-z_p}\sim L^{1/2}$. Numerical
     simulations (Fig. \ref{fig:rms_scale}) confirm the scaling forms in both the short and long time regimes.
     
     The same values of exponents were found for a puller on a fluctuating EW interface, in the $\omega = 1$ regime in \cite{cagnetta19} and a passive particle moving on the KPZ interface \cite{chin02,manoj04,nagar06}.


    \section{Puller on a fluctuating 1D interface}
     This regime refers to the hatched line in the dynamical parameter space (Fig. \ref{fig:1p_phasespace}), i.e., $u=1$ and $\beta_o<0$, and it corresponds to the regime addressed in \cite{cagnetta19} where the interface to particle update ratio, $\omega$ is 1. Here the puller is coupled to a fluctuating interface such that the dynamics of the interface at sites other than that of the puller belongs to EW class, while at the site of the particle the interface is preferentially being pulled up.
     
\begin{figure}[ht]
        \centering  
        \includegraphics[scale=0.25]{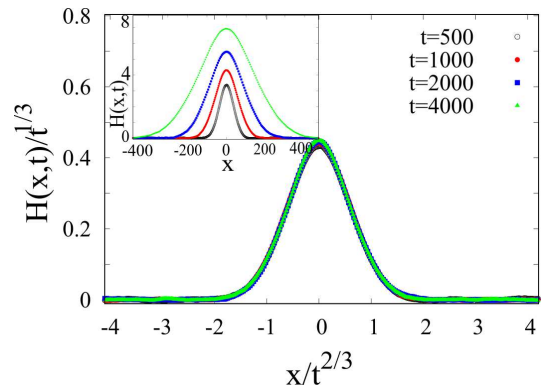}
        \caption{Scaled mean profile of a fluctuating interface coupled to a puller at different times. Unlike the tent-like pattern for the pulled static interface, here the peak at the origin is smooth.}
        \label{fig:profw1}
    \end{figure}

\begin{figure}[ht]
        \centering  
        \includegraphics[scale=0.25]{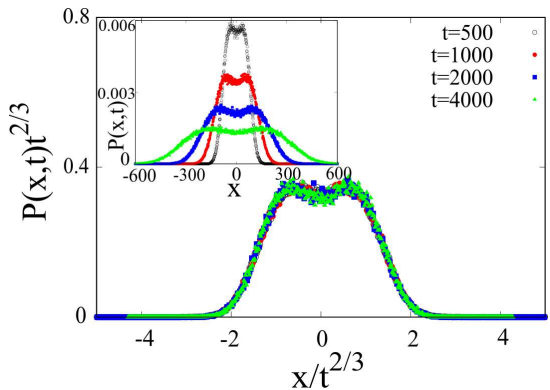}
        \caption{Scaled $P(x,t)$ for the puller on a fluctuating interface. The inset shows how $P(x,t)$ spreads with time. Note the slight dip at the origin. }
        \label{fig:probw1}
    \end{figure}

     We again observe a pattern in the mean interface profile $H(x,t)$ but with a smooth maximum at the origin instead of a tent-top (Fig. \ref{fig:profw1}). Furthermore, the probability distribution $P(x,t)$ for the particle shows a smooth dip at the origin instead of a linear cusp (Fig. \ref{fig:probw1}). Thus the effect of fluctuations is to wash out the sharp features observed in the static case while retaining the qualitative features. 
     
     The analysis in \cite{cagnetta19} shows that the particle displays an early superdiffusive ($<x^2(t)> \sim t^{4/3})$ motion and obeys the same scaling collapse as in Eq. (\ref{p_scale}). The dynamic exponent for the particle is $3/2$, implying that the time taken by the particle to attain steady state, $\tau_L$, scales as $\sim L^{3/2}$. It is shown in \cite{cagnetta19} that the dynamic exponent of the interface is $2$, i.e., the relaxation time for a system of size $L$ goes as $\sim L^2$, coinciding with that for an EW interface. The puller has no significant effect on the evolution in the bulk of the interface, which is EW-like. Thus the fluctuating interface and particle have different dynamic exponents $z$, unlike the case of puller on a static interface, where $z=3/2$ characterizes both.
  
     Moreover, the early time growth of roughness follows a power-law behavior, $ct^{\beta}$, independent of the size of the system, unlike for the pulled static interface. This is by virtue of the growth process occurring uniformly all over the interface. Consequently, the prefactor $c$ in Eq. (\ref{early_prefactor}) is a constant independent of $L$. Therefore from Eq. (\ref{crossover}), we get $cL^{z\beta} = L^{\alpha}$, retrieving the scaling relation in Eq. (\ref{abz_rel}).


    \section{Puller in a macroscopic valley}
    
\begin{figure}[ht]
        \centering
        \includegraphics[scale=0.62]{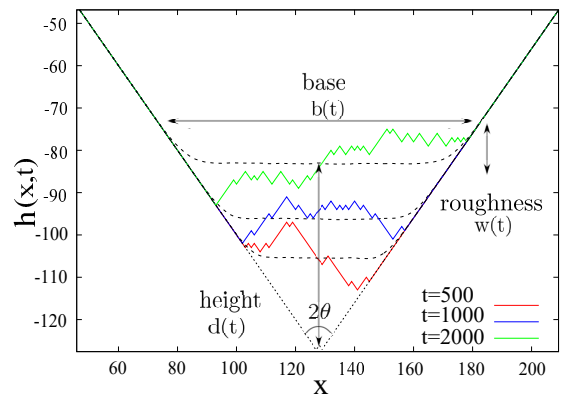}
        \caption{Starting from an interface in the form of a macroscopic triangular valley with a puller at the bottom, snapshots of the interface profiles are shown at different times. Colored curves represent the rough advancing front led by the puller, while the black curve shows the mean profile of the interface at that epoch. }
        \label{fig:typprof_time}
    \end{figure}

    In this section we impose an extreme, macroscopically different initial condition on the interface and study its effect on the interface motion and morphology. The interface initially is in the form of a giant valley obtained by setting $\tau_x = -1$ for $0\leqslant x<L/2$ and $\tau_x = 1$ for $L/2\leqslant x<L$. At $t=0$ the puller is placed at the bottom of the valley at $x=L/2$. We allow only $\vee \rightarrow \wedge$ flips at the site of the particle while the rest of the structure remains static. The puller sequentially overturns local valleys to hills, depleting the giant valley while riding on a rising rough front. The transverse motion of the puller is constrained by the boundary at the edge of the rough front, that itself evolves with time (Fig. \ref{fig:typprof_time}).
    When viewed upside down, the process can be  interpreted as a physical structure, like a mountain, being eroded by an active agent starting from the vertex.
    
    On average, the depleted portion of the macroscopic valley resembles a triangle of base $b$ and height $d$ (see Fig. \ref{fig:typprof_time}). The base of the triangle is the ensemble average of the rough front constructed and traversed by the active particle at that epoch. The front is defined by its mean height, $d(t)$ from the vertex of the macroscopic valley at $t=0$, its mean base length $b(t)$, and the fluctuations $w(t)$ about the mean height. In time the particle pulls the interface upwards, increasing $d$ as $\sim t^{\mu}$ while also expanding the horizontal extent as $b \sim t^{\lambda}$.
    
\begin{figure}[ht]
    \centering
    \includegraphics[scale=0.275]{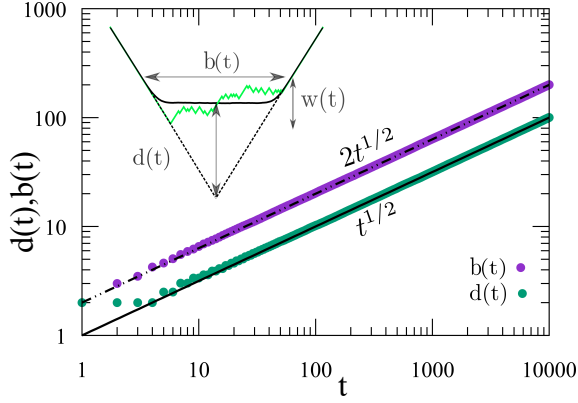}
    \caption{Growth of the average height $d$ and base $b$ with time.}
    \label{fig:base_height_growth}
    \end{figure}

    It is straightforward to deduce the exponents $\lambda$ and $\mu$ using the geometric constraints and conservation of area under the risen region, as invoked in Eq. (\ref{arealaw}). We observe from Fig. \ref{fig:typprof_time} that $2d/b = \tan{\theta} = 1$ at $t=0$ that implies $\lambda = \mu$. This condition holds true for the mean profile at each instant as $\theta$ remains fixed. As the number of valley-to-hill overturns and consequently the area under the risen region of the interface grows linearly with time, we also deduce, $\lambda + \mu = 1$. Therefore, $\lambda$ and $\mu$ both take the value $0.5$. The front gives an impression of advancing in two perpendicular directions dictated by the height and the base growth, both $d, b \propto t^{1/2}$ as shown in Fig. \ref{fig:base_height_growth}.

\begin{figure}[ht!]
       \centering
      \includegraphics[scale=0.25]{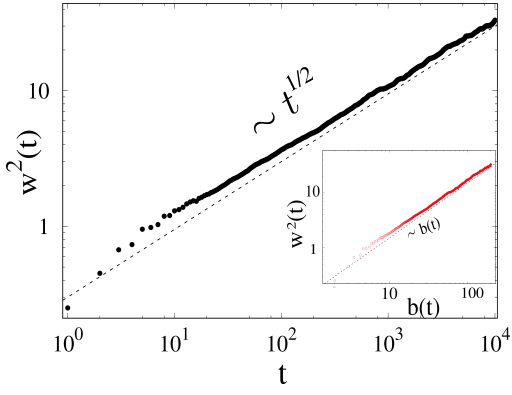}
      \caption{Mean-square roughness $w^2(t)$ of the front constructed by the puller in a macroscopic valley, growing with time. The inset shows that $w^2(t)$ is proportional to $b(t)$.}
       \label{fig:width_v_base_valley}
    \end{figure}

    A crucial difference that arises in this geometry is that the height-height correlations on the rough front are short-ranged, even though the system never attains steady state. The interface does not assume any distinctive mean pattern and rises upwards as a planar front. The mean square roughness of the front at time $t$ is defined as $w^2(t) = <\sum_i^{b}(h_i(t) -\sum_i^{b}(h_i(t)/b)^2>$. Interestingly, we see in Fig. \ref{fig:width_v_base_valley} that $w^2(t)$ grows as $\sim t^{1/2}$, proportional to the base length $b(t)$ at time $t$, a signature of uncorrelated slopes of the interface as in the steady state of the pulled static interface. The observation $w^2(t) \sim b(t)$ is suggestive that the front attains saturation roughness at each time epoch as it rises upwards.

    \section{Pusher on a fluctuating 1D interface}
    In this section, we analyze the effect of a pusher on a fluctuating interface. Recall the dynamics of a pusher; it slides down towards valleys and is more likely to convert hills into valleys than valleys into hills. The interface is a fluctuating EW interface where $u$ takes the value $1$. In the same vein as the puller on a fluctuating interface, the dynamics in the bulk of the interface belongs to the EW class whereas at the site of the particle it is locally KPZ-like as the interface is pushed downwards. A similar study for a pushing an active particle on a fluctuating interface was conducted in \cite{shauri2017a}
    
\begin{figure}[ht!]
    \centering
    \includegraphics[scale=0.675]{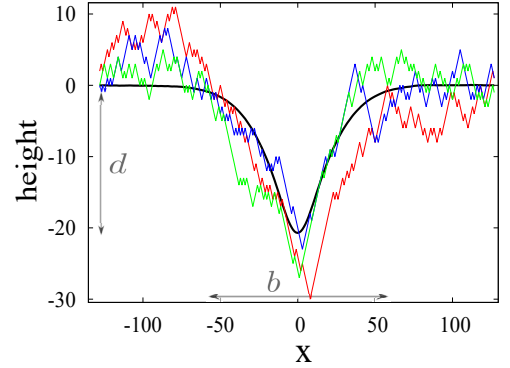}
    \caption{Typical profiles of a fluctuating interface coupled to a pusher (color), laid over the mean profile (black) averaged over many realization at t=1000. }
    \label{fig:typ_profile_w+ve}
    \end{figure}

    The pusher is placed at the origin at time $t=0$ on a ``flat" interface. As $t$ increases, the interface moves in the downwards direction and takes the form of a valley centered about the initial position of the pusher. Figure \ref{fig:typ_profile_w+ve} shows the typical profiles with respect to the mean profile, $H(x,t)$, averaged over several realizations.
    
    In a biological context, pushers in our model resemble the class of transmembrane proteins that impart protrusive forces of the cytoskeleton on the membrane, undulating it into finger like projections, instrumental in mechanosensing and cell migration \cite{bourne02,pollard03}. The valley-like deformation on the interface is comparable with filopodia, i.e., extended structures on the leading edge of a motile membrane.
    
    \begin{figure}[ht!]
    \centering
    \includegraphics[scale=0.315]{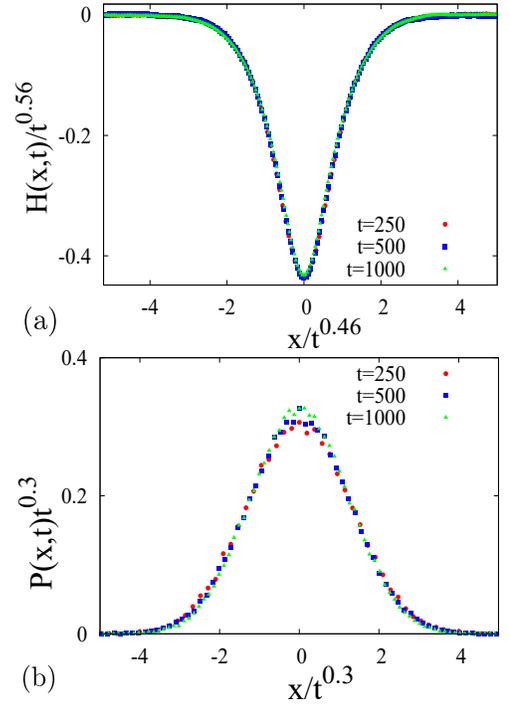}
    \caption{(a) Scaled mean profile in the pre-steady-state regime. (b) $P(x,t)$ of the particle as defined in the text. One can see that the scaling exponents are quite different.}
    \label{fig:scaled_w+ve}
    \end{figure}

    The base and height of $H(x,t)$ grow as $\sim t^{\mu'}$ and $\sim t^{\lambda'}$ respectively where $\mu' \simeq 0.56$ and $\lambda' \simeq 0.46$. Unlike the puller, here the base of the pattern grows slower than the height, that suggests an enhanced pushing activity and a restricted transverse motion. Both the average profile $H(x,t)$ and the particle position probability distribution $P(x,t)$ exhibit a scaling collapse for different times (Fig. \ref{fig:scaled_w+ve}). But the scaling exponent for the two observables does not coincide, suggesting a separation of time scale for the particle and interface dynamics.
    
\begin{figure}[ht!]
     \hspace{-0.5cm}
     \includegraphics[scale=0.4]{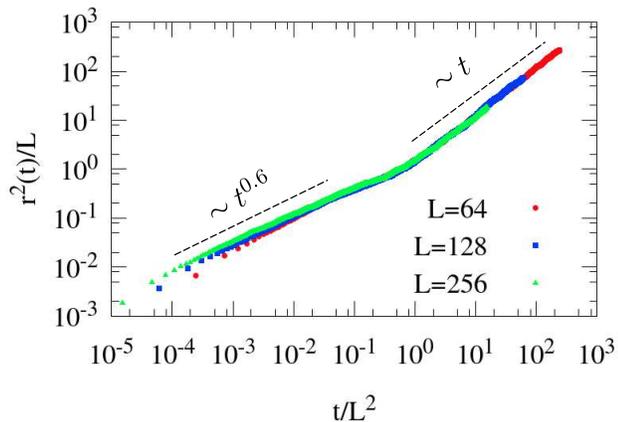}
     \caption{Mean-square displacement for a pusher on a fluctuating interface. Scaling with different sizes was seen with the dynamic exponent $z'  = 2$ as the best fit. The dashed lines delineate the early subdiffusive $r^2(t) \sim t^{2\lambda'}$ and later diffusive $r^2(t) \sim t$ regime. }
     \label{fig:msd_w+ve}
     \end{figure}

     \noindent \textit{Particle motion}: Figure \ref{fig:msd_w+ve} displays the RMS displacement, $r(t) = \sqrt{<x^2(t)>}$, of the pusher. The initial slow subdiffusive regime where the RMSD grows as $r(t) \sim t^{\beta'}$, with $\beta' \simeq 0.3$, can be attributed to the ``digging" activity of the particle, that confines the particle motion to a valley. The motion of the pusher is dictated by $P(x,t)$ as the transverse length scale in $P(x,t)$ grows as $\sim t^{\beta'}$. The transition to a diffusive regime occurs at $\tau \sim L^{z'}$ where the value $z'$ is observed to be close to 2. 
    
    The differences in the scaling properties of $H(x,t)$ and $P(x,t)$ can be understood as an outcome of the separation of length scales between particle motion and the consequent response of the interface. The scaling of $P(x,t)$ is solely a function of the particle motion; whether puller or pusher. As for the average profile, it was shown in \cite{shauri2017a}, that a fluctuating Edwards-Wilkinson interface assumes a mean structure about the particle with a diffusive growth law i.e. $H(x,t) \sim t^{1/2}$ around a static pusher, implying that the interface deforms with a length scale that grows as $\sim t^{1/2}$. Comparing the displacement of a puller and pusher on a fluctuating interface, we have established that the puller performs a superdiffusive motion $r(t) \sim t^{\beta}$ whereas the pusher displays a subdiffusive motion $r(t) \sim t^{\beta'}$, where $\beta$ and $\beta'$ are $\simeq 0.66$ and $\simeq 0.3$ respectively. Clearly, the motion of the puller is faster than the growth of the response length scale of the underlying fluctuating interface ($t^{\beta}>t^{1/2}$), whereas it is slower for the pusher  ($t^{\beta'}<t^{1/2}$). Thus, the effective scaling exponent manifested in the profile, which is dictated by the faster growing length scale in the system, follows $\sim t^{2/3}$ for a puller and $\sim t^{1/2}$ for a pusher on a fluctuating interface.


    \section{Puller on a static 2D interface}

    In this section, we study the dynamics of a puller on a static 2D interface. As discussed in Sec. \ref{model}, the puller increments the height of its current site by $+2$ only if it is lower than all of its neighboring sites. Once on top of a hill, the particle randomly chooses a site out of its four neighbours and slides down if the chosen site is lower in height. Initially, the puller is placed at the center of the 2D grid, designated as the origin.
    
\begin{figure}[ht!]
    \centering
    \includegraphics[scale=0.33]{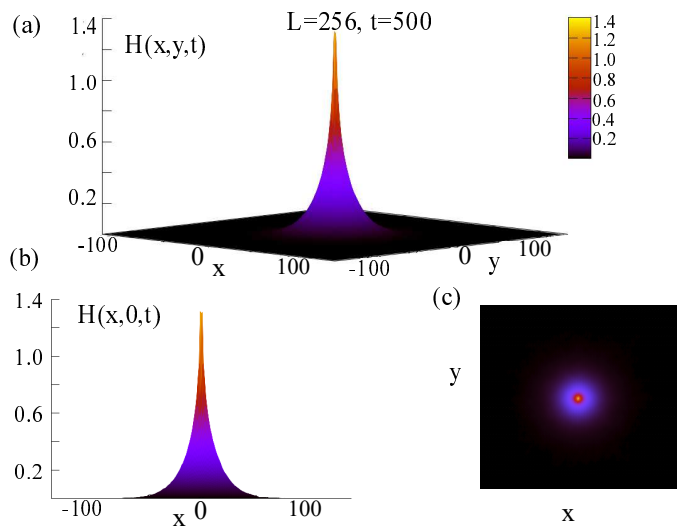}
    \caption{The ensemble average height profile (a) on a 2D interface, (b) projection in the $H-x$ plane, and (c) top view of the profile.}
    \label{fig:2dprofile}
    \end{figure}

    \noindent We see a mean profile $H(x,y,t)$, akin to the mean interface pattern in one dimension, emerging on the interface (Fig. \ref{fig:2dprofile}). As in one dimension, the profile has a tent-like structure with a linear cusp at the origin. It is observed that the pattern is circularly symmetric about the origin, i.e., $H(x,y,t) = f(r,t)$ where $r = (\sqrt{x^2 + y^2})$.

\begin{figure}[ht!]
    \centering
    \includegraphics[scale=0.6]{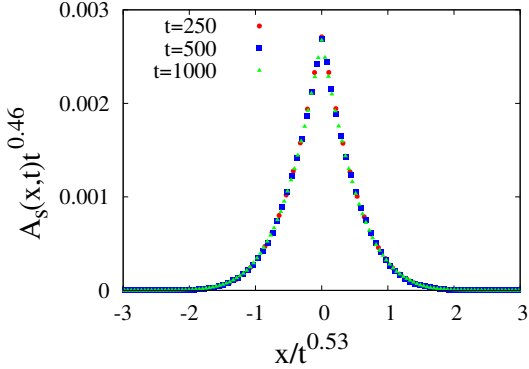}
    \caption{The scaled differential area $A_s$ at each $x$ as seen from the $y=0$ plane. The area under the curve represents the volume under the mean profile.}
    \label{fig:2dscaled_prof}
    \end{figure}

   As in one dimension, aspects of scaling are seen in two dimension as well. Let us define a quantity $A_s$ as the area of a differential strip of unit breadth and length $L$ along, say, the $x$ axis, as a representative direction. Therefore, we have, $A_s(x,t) = \sum_{y=0}^{L} h(x,y,t)/L$. We observe that it follows the scaling form:
    \begin{equation}
        A_s(x,t) \approx t^{\mu''}\mathcal{A}(\frac{x}{t^{\lambda''}})
        \label{diffarea}
    \end{equation}where $\lambda'' \simeq 0.53$ and $\mu'' \simeq 0.46$, implying the area of the projected mean profile grows as $\sim t^{\mu''}$ while its base length $b$ grows as $\sim t^{\lambda''}$ (Fig. \ref{fig:2dscaled_prof}). This relation is robust with respect to the choice of direction. Also in two dimensions, the argument bearing out Eq. (\ref{arealaw}) holds true: the number of valley to hill overturns grows linearly with time. This translates into the volume of the risen region on the 2D membrane growing proportional to $t$.  The volume here is evaluated as $\sim A_s \times b \sim t^{\mu''+\lambda''}$ and we see $\lambda''+\mu'' = 1$. The characteristic length scale in two dimensions grows as $\sim t^{\lambda''}$ with $\lambda'' = 0.53$.\\

    \begin{figure}[ht]
    	\centering
    	\includegraphics[scale=0.33]{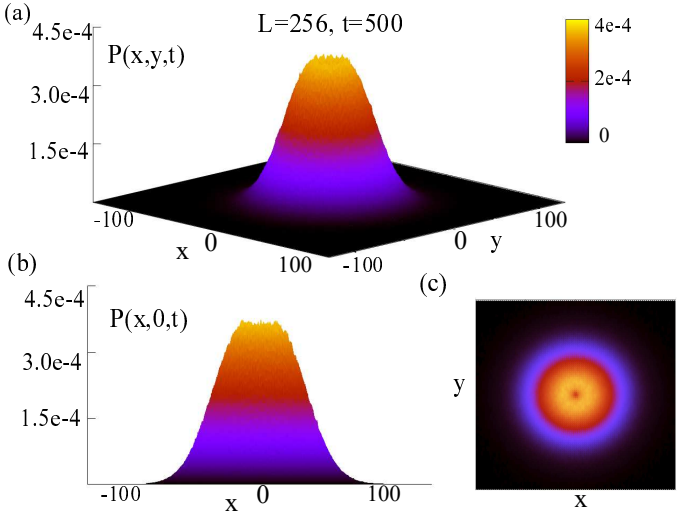}
    	\caption{The particle probability distribution $P(x,y,t)$ (a) on a 2D interface, (b) projection in the $P-x$ plane, and (c) top view of the distribution.}
    	\label{fig:2dprob}
    \end{figure}

    We also study the probability distribution of the particle position, $P(x,y,t)$. Like $H(x,y,t)$, $P(x,y,t)$ is symmetric about the origin (Fig. \ref{fig:2dprob}). We observe a local minimum at the origin with a linear cusp similar to that in the 1D case. We can deduce the scaling in probability by employing the following normalization condition:
    
    \begin{equation}
        \int_0^L\int_0^L P(x,y,t)dxdy = 1
        \label{norm}
    \end{equation}
    \noindent We know from Eq. \ref{diffarea}, that the characteristic length scale and hence both $x$ and $y$ grow as $\sim t^{\lambda''}$. Assuming that $P(x,y,t)$ grows in a power law fashion as $\sim t^{\gamma''}$, we rewrite Eq. (\ref{norm}) as:
    \begin{equation}
        c\int_0^{\tau'}\int_0^{\tau}t^{\gamma''}t^{\lambda''-1}\tau^{\lambda''-1}dtd\tau = 1
    \end{equation}
    
\begin{figure}[ht!]
    \centering
    \includegraphics[scale=0.6]{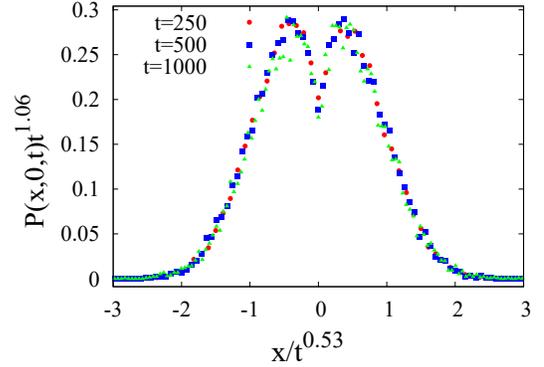}
    \caption{A cross-section of scaled $P(x,y,t)$ along the $y=0$ plane. Scaling holds for the cross section of $P(x,y,t)$ along any plane passing through $x=0,y=0$. }
    \label{fig:2dscaled_prob}
    \end{figure}

    \noindent Evaluating this integral and comparing the power of $\tau'$ on both sides we derive the condition: $\gamma'' + 2\lambda'' = 0$. Therefore, $\gamma'' = -2\lambda''$ i.e. the probability $P(x,y,t)$ scales as $\sim t^{-2\lambda''}$. We observe that the cross section of $P(x,y,t)$ along the $y=0$ plane obeys the following scaling collapse (Fig. \ref{fig:2dscaled_prob}):
    
    \begin{equation}
        P(x,0,t) \approx t^{-2\lambda''}\mathcal{P}(\frac{x}{t^{\lambda''}})
        \label{2d_prob_scaling}
    \end{equation}

\begin{figure}[ht!]
   \centering
   \includegraphics[scale=0.235]{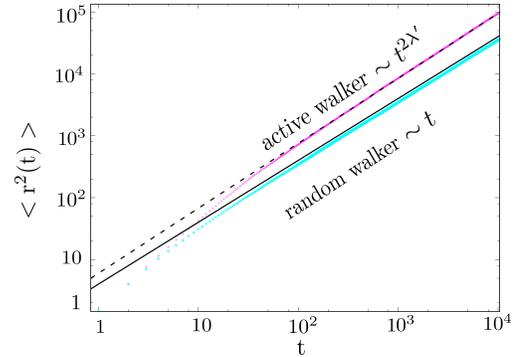}
   \caption{In two dimensions, the rms displacement for the active puller is close to diffusive. }
   \label{fig:2dstats}
   \end{figure}

    \noindent Equation (\ref{2d_prob_scaling}) implies that the lateral mean square displacement of the particle grows as $<r^2(t)> \sim t^{2\lambda''}$ which is what we observe in Fig. \ref{fig:2dstats}. The motion of the puller in two dimensions is close to diffusive, in contrast to its superdiffusive motion in one dimension. The slower motion of the puller in two dimensions is reflected in a larger time scale $L^{z''}$ for the system to achieve steady state where $z'' = 1/\lambda'' \simeq 1.9$ as opposed to $z=1.5$ in one dimension.

    \section{Conclusion}
    
    We have discussed the response of an interface to a single active particle that moves on it stochastically, and is capable of deforming it by searching for local valleys (for a puller) or local hills (for a pusher), and
    reconfiguring them. Our work builds on previous lattice model studies of particle-interface interactions \cite{cagnetta19} and extends the range of study by tuning the nature of the particle activity (pushing or pulling) and the degree of fluctuations intrinsic to the interface.
    
    The most interesting case is that of a puller on an otherwise static interface where the particle actively overturns valleys to hills, then slides down the slopes in search of fresh valleys to overturn. This overturn-slide-search sequence ensures that the system is always in a state of activity. Unlike the traditional studies of passive particles driven by a stochastically evolving landscape \cite{drossel00,nagar06,chin02,manoj04,singha18}, in our case the driving is done by the active particle and the driven component is the interface. In the process, the interface acquires a small velocity and assumes an interesting mean profile $H(x,t)$. At the same time, the probability distribution of the particle location $P(x,t)$ has an interesting double-peaked form. Both $H(x,t)$ and $P(x,t)$ exhibit scaling in space and time. Thus, we have a model for an interface that, on its own is frozen and at rest, but it can be driven when interacting with an active agent that traverses it and pulls it up, giving the interface a distinct morphology. This may be a useful prototype in the biological context of a cell membrane that develops protrusions and acquires motility in response to an external cue, caused by protein assembly at the moving front \cite{pollard03}.
    
    A significant point about the dynamics is the breakdown of customary scaling for the width $w$ of the interface. The observed values of dynamic exponent $z$, growth exponent $\beta$, and roughness exponent $\alpha$ violate the scaling condition $z = \alpha/\beta$, which links together the early time and steady-state growth of width, familiar from many interface growth processes \cite{ew,barabasi,fam_vic,kpz}. This breakdown was shown to be an outcome of an $L$-dependent factor $\sim 1/L$ in the initial growth of width, coming from the fact that the growth is a very local process. Interestingly, we were able to deduce the exponent values on the basis of simple arguments. For the case of a pulled fluctuating interface with dynamics belonging to the Edwards-Wilkinson class, by contrast, the evolution is dominated by the EW dynamics occurring at $L-1$ sites, and the width follows normal scaling relations \cite{cagnetta19}.
    This raises an interesting general question regarding interface growth, when interface evolution occurs at a sublinear number of sites. If this leads to the width growing as $w^2 \sim L^{\sigma-1}t^{2\beta}$, (where $\sigma < 1$), then a generalization of the argument in Eq. (\ref{crossover}) leads to the revised scaling relation of the form: $2z\beta + \sigma - 1 = 2\alpha$. For customary interface growth processes, $\sigma$ is 1, while in our study $\sigma$ is $0$. It would be interesting to verify the generalized scaling relation in cases in which $\sigma$ assumes other values.

    We also studied the effect of changing the initial condition away from a flat interface, allowing for intrinsic fluctuations of the interface, having the active particle be a pusher rather than a puller, and extending the static interface study to two dimensions. In every case, we found that the mean profile and the probability distribution of particle displacement exhibit features of scaling.

    We conclude that a single active particle driving an interface gives rise to a remarkably rich set of interconnected phenomena. Expanding the study to include a macroscopic number of pushers and pullers would be necessary to simulate a picture closer to the highly non-equilibrium environment of the plasma membrane, with an emphasis on cooperative effects among active agents and the consequent membrane deformations.


    \section{Acknowledgements}
    This work forms part of a thesis submitted by PB in partial fulfillment of the requirement for the degree of Master of Science awarded by Indian Institute of Space Science and Technology, Thiruvananthapuram. PB would like to thank TIFR Hyderabad for hospitality and for the opportunity to work in a scientifically encouraging environment. We would like to thank F. Cagnetta, M. R. Evans, Pushpita Ghosh, Tapas Singha, Tamal Das, and Kabir Ramola for useful discussions.


\begin{thebibliography}{10}
   	
   	\bibitem{activematter}
   	M. C. Marchetti, J. F. Joanny, S. Ramaswamy, T. B. Liverpool, J. Prost, M. Rao and R. A. Simha, 
   	\newblock Rev. Mod. Phys. {\bf 85}, 1143 (2013).	
   	
   	
   	\bibitem{prost96}
   	J.  Prost  and  R.  Bruinsma, 
   	\newblock Europhys.  Lett. {\bf 33}, 321 (1996).	
   	
   	
   	\bibitem{madan01}
   	S. Ramaswamy and M. Rao, 
   	\newblock C. R. Acad. Sci. Paris, S\'erie IV {\bf 2}, 817 (2001).
   	
    S. Ramaswamy and M. Rao, 
   
   	\bibitem{act2}
   	S. Ramaswamy, J. Toner and J. Prost, 
   	\newblock Phys. Rev. Lett. {\bf 84}, 3494 (2000).	
   	
   	
   	\bibitem{fluidmos}
   	S. J. Singer and G. L. Nicolson, 
   	\newblock Science {\bf 175}, 720 (1972).
   	
   	\bibitem{bourne02}
   	H. R. Bourne and O. Weiner, 
   	\newblock Nature {\bf 419}, 21 (2002).
        
    \bibitem{pollard03}
    T.D. Pollard and  G.G. Borisy, 
    \newblock Cell {\bf 112}, 453 (2003).
   	
   	\bibitem{gov18}
   	N. Gov, 
   	\newblock Phil. Trans. R. Soc. B {\bf 373}, 20170115 (2018).
   	   	
   	
   	\bibitem{howard}
   	J. Howard , S.W. Grill, J.S. Bois, 
   	\newblock Nat. Rev. Mol. Cell Biol. {\bf 12}, 6 (2011).
   	   	
    
    \bibitem{priezzhev96}
    V. B. Priezzhev, D. Dhar, A. Dhar, and S. Krishnamurthy,
    \newblock Phys. Rev. Lett. {\bf 77}, 5079 (1996).
        	
	
	\bibitem{cagnetta19}
	F. Cagnetta, M. R. Evans and D. Marenduzzo, 
	\newblock Phys. Rev. E {\bf 99}, 042124 (2019).
   		   	   
   
   \bibitem{veksler07}
   A. Veksler and N. Gov, 
   \newblock Biophys. J. {\bf 93}, 3798 (2007).	
   
   \bibitem{gopi06}
   N. Gov and A. Gopinathan, 
   \newblock Biophys. J. {\bf 90}, 2 (2006).	
      

   \bibitem{unpub}
   P. Bisht and M. Barma, Unpublished.   
   
   
   \bibitem{shauri2017a}
   S. Chakraborty, S. Chatterjee and M. Barma, 
   \newblock Phys. Rev. E {\bf 96}, 022127 (2017).
   
   
   \bibitem{shauri2017b}
   S. Chakraborty, S. Chatterjee and M. Barma, 
   \newblock Phys. Rev. E {\bf 96}, 022128 (2017).
   	
    
    \bibitem{cagnetta18}
    F. Cagnetta, M. R. Evans and D. Marenduzzo, 
    \newblock Phys. Rev. Lett. {\bf 120}, 258001 (2018).
     
    
    \bibitem{ew}
    S. F. Edwards and D. R. Wilkinson, 
    \newblock Proc. R. Soc. London, Ser. A {\bf 381}, 17 (1982).
        
    
   	\bibitem{chin02}
   	C.-S. Chin, 
   	\newblock Phys. Rev. E {\bf 66}, 021104 (2002).
   	   	
   	
   	\bibitem{drossel00}
   	B. Drossel and M. Kardar, 
   	\newblock Phys. Rev. B {\bf 66}, 195414 (2002).
   	   	
   	\bibitem{manoj04}
   	M. Gopalakrishnan,
   	\newblock  Phy. Rev. E {\bf 69}, 011105 (2004).
   	
   	
   	\bibitem{nagar06}
   	A. Nagar, S. N. Majumdar and M. Barma, 
   	\newblock Phys. Rev. E {\bf 74}, 021124 (2006).
   	  	   	
   	
   	\bibitem{singha18}
   	T. Singha and M. Barma, 
   	\newblock Phys. Rev. E {\bf 98}, 052148 (2018).
   	   	
   	\bibitem{meakin}
   	P. Meakin,
   	\newblock \textit{Fractals, Scaling and Growth far from Equilibrium}, (Cambridge University Press, Cambridge, England, 1998).
   	
   	
   	\bibitem{manoj2d}
   	G. Manoj and M. Barma, 
   	\newblock J. Stat. Phys. {\bf 110}, 3 (2003).
   	
   	
   	\bibitem{barabasi}
    A. L. Barab\'asi and H. E. Stanley,
   	\newblock\textit{Fractal Concepts in Surface Growth}, (Cambridge University Press, Cambridge, England, 1995).
   	
   	
   	\bibitem{fam_vic}
   	F. Family and T. Vicsek, 
   	\newblock J. Phys. A {\bf 18}, L75 (1985).
   	
   	
	\bibitem{kpz}
	M. Kardar, G. Parisi and Y.-C. Zhang,
	\newblock Phys. Rev. Lett. {\bf 56}, 889 (1986).
	
	 
   \end{thebibliography}
\end{document}